\documentstyle[preprint,aps,rotate,epsfig]{revtex}
\tighten
\begin{document}
\title{Baryon flow from SIS to AGS energies}
\author{P. K. Sahu$^1$\thanks{JSPS Research Fellow.},
W. Cassing$^2$, U. Mosel$^2$ and A. Ohnishi$^1$
}
\address{
$^1$ Division of Physics, Graduate School of Science, Hokkaido University,
Sapporo 060-0810, Japan\\
$^2$ Institut f\"ur Theoretische Physik, Universit\"at Giessen,
D-35392 Giessen, Germany
}

\maketitle

\begin{abstract}

We analyze the baryon sideward and elliptic flow from SIS (0.25
$\sim$ 2 $A$GeV) to AGS (2 $\sim $11.0$A$GeV) energies for Au + Au
collisions in the relativistic transport model RBUU that includes all
baryon resonances up to a mass of 2 GeV as well as string degrees of
freedom for the higher mass continuum. There are two
factors which dominantly determine the baryon flow at these
energies: the momentum dependence of the scalar and vector
potentials and the resonance-string degrees of freedom. We fix the
explicit momentum dependence of the nucleon-meson couplings within
the NL3 parameter set by the nucleon optical potential up to 1
GeV of kinetic energy. When assuming the
optical potential to vanish identically for
$E_{kin} \geq 3.5$ GeV we simultaneously reproduce the sideward
flow data of the FOPI, EOS, E895 and E877 collaborations, the
elliptic flow data of the EOS, E895 and E877 collaborations,
and approximately the rapidity and transverse mass distribution
of protons at AGS energies.
The gradual change from hadronic to string degrees of freedom
with increasing bombarding energy can be viewed as a transition
from {\it hadronic} to {\it string} matter, i.e. a
dissolution of hadrons.

\end{abstract}



\newpage

\section{Introduction}
%
The nuclear equation of state (EoS) plays a central role in
heavy-ion collisions \cite{cass99} -- \cite{sorge97} as well as
for the maximum mass of neutron stars and supernova explosions.
Therefore, the nuclear EoS is one of the most challenging topics
in nuclear physics, i.e. to understand the nature of the nuclear
force at high density and/or temperature. In nucleus-nucleus
collisions the transverse flow observables, including directed
\cite{rei97} and elliptic flow \cite{dani98,sorge97}, as well as
subthreshold particle production\cite{Cass} are sensitive to the
nuclear EoS. Recently, both the directed transverse flow (sideward
flow) and the flow tensor (elliptic flow) have been measured and
reported in Refs. \cite{FOPI} -- \cite{EOS}, \cite{aE877} --
\cite{bE895} for heavy-ion (Au + Au) collisions in the incident
energy range of $1 A$ GeV $\le E_{inc} \le 11 A$ GeV. In this
energy range the directed transverse flow first grows, saturates
at around 2 $A$ GeV and then decreases experimentally with energy
showing no minimum as expected from hydrodynamical calculations
including a first order phase transition in the EoS
\cite{Rischke}. On the other hand the elliptic flow  measured
experimentally changes its sign from negative (squeeze-out) to
positive (in-plane) as a function of incident energy in the range
of $1 A$ GeV $\le E_{inc} \le 11 A$ GeV.

In this work we analyze the directed transverse flow
in the incident energy range starting from GSI-SIS to the BNL-AGS regime
by using the relativistic transport model RBUU
which includes the two essential ingredients, i.e.
i) the momentum-dependent potentials and ii) the
resonance/string degrees of freedom.
With these ingredients fixed we examine the sideward and the elliptic flow
of protons as a function of the
incident energy up to AGS energies of $\sim 1 - 11 A$ GeV.

We recall that relativistic transport models have been extensively used
to describe the heavy-ion data at energies starting from
the SIS at GSI to the SPS regime at CERN,
\cite{cass99,dani98,Dani,mar94,ehe96,bas97,li97}.
Among them the Relativistic Boltzmann-Uehling-Uhlenbeck (RBUU)
approach is one of the most successful models.
It incorporates the relativistic mean-field (RMF) theory,
which is applicable to various nuclear structure problems as well
as for neutron star studies~\cite{SahuNS}.
Thus it is possible to refine the mean-field part of RBUU
by incorporating constraints from the latter fields.
Here we base our study on the NL3 parameter set from Ref. \cite{lan91}
since this parameter set has been widely applied in the analysis
of heavy-ion collisions \cite{sahu98,bli99,soff99}.

The most simple versions of RMF theories
assume that the scalar and vector fields
are represented by point-like meson-baryon couplings.
These couplings lead to a linearly growing
Schr{\"o}dinger-equivalent potential
in nuclear matter as a function of the kinetic energy $E_{kin}$,
which naturally explains the energy dependence of the nucleon optical potential
at low energies ($\leq 200$ MeV).
However, a simple RMF does not describe the nucleon optical potential
at higher energies, where the optical potential deviates substantially
from a linear function and saturates at $E_{kin} \approx$ 1 GeV.
Since the energy dependence of sideward flow is controlled in part
by the nucleon optical potential, the simple RMF
cannot be applied to high-energy heavy-ion collision problems.
In order to remedy this aspect, some RBUU approaches
invoke an explicit momentum dependence of the coupling constant, i.e.
a form factor for the meson-baryon couplings \cite{cass99,ehe96}.

In our earlier work on directed flow in nucleus-nucleus collisions
\cite{sahu98} we showed that the scalar and vector self energies
for nucleons including a momentum and density dependence are the key quantities
which determine the behaviour of flow at SIS energies
and explain the kinetic energy dependence of the nucleon optical potential
as well.      In this study
we proceed with the systematic analysis of flow up to AGS energies using
the latter explicitly momentum-dependent relativistic mean fields.

A further important ingredient at AGS energies are the
resonance/string degrees of freedom which are excited during the
reaction in high energy baryon-baryon or meson-baryon collisions.
While at SIS energies particle production mainly occurs through
baryon resonance production and their decay, the string
phenomenology is found to work well at SPS energies \cite{cass99}.
We note here that both mechanisms really describe the same
physics, i.e. the decay of excited baryons or mesons. The only
essential difference is the treatment of the decay of these
excited states: while the baryon resonances all decay to $N$ +
meson (or $N^*$ + meson) the strings can decay to $N n \pi (n >
1)$. Due to these different decay schemes there are various ways
to implement elementary cross sections in transport
models~\cite{cass99,dani98,bli99,soff99,nara99,hom98}. One of the
extremes is to parameterize all possible cross sections for
multi-pion production, $NN \to NN n\pi (n \ge 3)$ only through $N,
\Delta, \pi$ degrees of freedom; the other extreme is to fully
apply string phenomenology in this energy region without employing
any resonances. Although it is possible to reproduce the
elementary cross sections  from $NN$ and $\pi N$ collisions and
the inclusive final hadron spectra in heavy-ion collisions within
these different models, we expect that differences should appear
in the dynamical evolution of the system, e.g. in the
thermodynamical properties \cite{hom98,nara97} and in collective
flow~\cite{hom98}. For example, if thermal equilibrium is achieved
at a given energy density, models with a larger number of degrees
of freedom including strings will give smaller temperature and
pressure. Since the transverse flow is partly made up from this
pressure, we expect a reduction of flow for models with a larger
number of degrees of freedom (strings) while a pure resonance gas
including high mass baryon excitations may be of higher
temperature and develop a larger pressure \cite{Frank1}. In order
to investigate the sensitivity of observables to this point we
introduce an energy scale $\sqrt{s_{sw}}$ that separates the two
production mechanisms.

We have employed here a new transport code that incorporates all
nucleon resonances up to a mass of 2 GeV \cite{effe,man92} and the
Lund string model \cite{lund} for higher excitations as in the
Hadron-String-Dynamics (HSD) approach \cite{cass99,ehe96}. In the
practical implementation for $NN$ or $MN$ collisions at invariant
energies lower (higher) than $\sqrt{s_{sw}}$ resonances (strings)
are assumed to be excited. In view of our basis space of baryon
resonances up to 2 GeV the transition energy $\sqrt{s_{sw}}$ has
to be below 4 GeV. Since the transverse pressure from strings is
smaller than that from resonances the transverse mass spectra of
baryons should become softer for smaller $\sqrt{s_{sw}}$. As
described below we fix this parameter by $\sqrt{s_{sw}} \approx
3.5$ GeV in fitting the transverse mass spectra of protons in
central Au + Au collisions at AGS energies.

We organize our work as follows: In Section 2 we briefly describe
the relativistic transport approach with known constraints on the
momentum dependence of the scalar and vector self energies. In
Section 3 we will systematically study the transverse mass spectra
for protons with and without momentum-dependent
potentials as well as resonance/string degrees of freedom and
compare the calculated sideward and elliptic flow with the
experimental data. Section 4 concludes this study with a summary
and discussion of open questions.

\section{The relativistic transport model}
In the present calculation we perform a theoretical analysis along
the line of the relativistic transport approach RBUU which is
based on a coupled set of covariant transport equations for the
phase-space distributions $f_{h} (x,p)$ of a hadron $h$
\cite{cass99,ehe96,web93}. The model inputs are the nuclear mean
fields and the (in-medium) elementary hadron-hadron cross
sections. In the relativistic transport approach the nuclear mean
field contains both  vector- and scalar-potentials $U^\nu$ and
$U^S$, respectively, that depend on the nuclear density and
momentum. In this work, these mean fields are calculated on the
basis of the same Lagrangian density as considered in our earlier
calculations \cite{sahu98}, which contain nucleon, $\sigma$ and
$\omega$ meson fields and nonlinear self-interactions of the
scalar field (cf. NL3 parameter set~\cite{lan91}). The scalar and
vector form factors at the vertices are taken into account in the
form \cite{ehe96}
\begin{equation}
\label{form}
    f_s({\bf p})=\frac{\Lambda_s^2-\frac{1}{2}{\bf p}^2}{\Lambda_s^2+{\bf p}^2}
    \qquad\mbox{and}\qquad
    f_v({\bf p})=\frac{\Lambda_v^2-\frac{1}{6}{\bf p}^2}{\Lambda_v^2+{\bf p}^2}\ ,
\end{equation}
where the cut-off parameters $\Lambda_s = 1.0$ GeV
and $\Lambda_v = 0.9$ GeV
are obtained by fitting the Schr\"odinger equivalent potential,
\begin{equation}
\label{pot}
U_{sep} (E_{kin}) = U_s + U_0 + \frac{1}{2M} (U_s^2-U_0^2) +
\frac{U_0}{M} E_{kin},
\end{equation}
to Dirac phenomenology for intermediate energy proton-nucleus
scattering \cite{ham90}. The above momentum dependence is computed
self-consistently on the mean-field level; in the actual
calculations we evaluate $U^0$ and $U^S$  in the local rest frame
of the surrounding nuclear matter and then perform a Lorentz
transformation to get $U^\mu$ in the calculation frame. Thus
neglecting a nonlocality in time, this evaluation of the potential
is practically covariant.

The resulting Schr\"odinger equivalent potential (\ref{pot})
at density $\rho_0$ is shown in Fig.~1 as a
function of the nucleon kinetic energy
with respect to the nuclear matter at rest in comparison
to the data from Hama et al. \cite{ham90}.
The increase of the Schr\"odinger equivalent potential up to
$E_{kin} = 1$ GeV is decribed quite well;
then the potential decreases and is set to zero above 3.5 GeV.

For the transition rate in the collision term of the transport
model we employ in-medium cross sections as in Ref. \cite{effe}
that are parameterized in line with the corresponding experimental
data for $\sqrt{s} \leq \sqrt{s_{sw}}$. For higher invariant
collision energies we adopt the Lund string formation and
fragmentation model \cite{lund} as incorporated in the HSD
transport approach \cite{cass99,ehe96} which has been used
extensively for the description of particle production in
nucleus-nucleus collisions from SIS to SPS energies \cite{cass99}.
One might speculate that the transition rates change substantially
with increasing baryon density. However, in this case the overall
reproduction of baryon and pion rapidity distributions from SIS to
SPS energies would be spoiled. In Ref. \cite{cass99} no indication
was found for the light quark sector whereas the enhanced
production of strangeness at AGS energies (found experimentally)
might be attributed to enhanced transition rates. Since this is
presently an open question and discussed in a controversal manner,
we discard the physics of strange hadrons in this work and
concentrate on the collective aspects of nucleons and pions.

In the present relativistic transport approach (RBUU) as in our
earlier work \cite{sahu98} we ex\-pli\-cit\-ly propagate nucleons
and $\Delta$'s as well as all baryon resonances up to a mass of 2
GeV with their isospin degrees of freedom \cite{effe,man92,hom96}.
Furthermore, $\pi, \eta$, $\rho$, $\omega$, $K, \bar{K}$ and
$\sigma$ mesons are propagated, too, where the $\sigma$ is a short
lived effective resonance that describes $s$-wave $\pi \pi$
scattering. For more details we refer the reader to
Refs.~\cite{effe,hom96} concerning the low energy cross sections
and to Refs. \cite{cass99,ehe96} with respect to the
implementation of the string dynamics, respectively. We note that
throughout this work we have neglected the momentum dependence of
the potentials in the collision term in order to save
computational time for the systematic analysis.

\section{Comparison to experimental data}

\subsection{Transverse mass spectra of protons}
In Fig.~2  we show the proton transverse
mass spectra (a) and rapidity distribution (b)
in a central collision of Au + Au at 11.6 $A$ GeV ($b$ $< 3.5$ fm)
for $\sqrt{s_{sw}}$ = 2.6 GeV (dotted histograms) and 3.5 GeV (solid
histograms) in comparison
to the experimental data of the E802 collaboration \cite{E802}. A cascade
calculation (crosses) is shown additionally for $\sqrt{s_{sw}}$ = 3.5 GeV
to demonstrate the effect of the mean-field potentials which lead to a
reduction of the transverse mass spectra  below 0.3 GeV and a substantial
hardening of the spectra.
As expected, the transverse mass spectrum is softer
for smaller $\sqrt{s_{sw}}$ due to the larger number of
degrees of freedom in the string model relative to the resonance model.

We note that strings may be regarded as hadronic excitations in
the continuum of lifetime $t_F \approx$ 0.8 fm/c (in their rest frame)
that take over a significant part of the incident collision energy
by their invariant mass. They decay dominantly to light baryons
and mesons and only to a low extent to heavy baryon resonances.
Thus the  number of particles for fixed system time is larger for
string excitations than for the resonance model; in the former several
hadrons propagate as a single heavy resonance which might be
regarded as a cluster of a nucleon + $n$ pions. As a consequence
the translational energies are suppressed in string excitations
and, as a result, the temperature as well as the pressure are
smaller when exciting strings.

From the above comparison with the experimental transverse mass
spectra for protons in Fig. 2 we fix $\sqrt{s_{sw}} \approx$ 3.5 GeV; this
implies that binary final baryon channels dominate in our transport model
up to $\sqrt{s} \approx$ 3.5 GeV which corresponds to a proton
laboratory energy $T_{lab}$ of about 4.6 GeV. We mention that $\sqrt{s_{sw}}$
may be changed by $\pm$ 0.3 GeV and still describe the proton spectra in
Fig. 2.

\subsection{Directed flow for Au + Au collisions}

We now turn to collective flow. The calculations are performed for
the impact parameter $b=6 fm$ for Au + Au systems, since for this
impact parameter we get the maximum flow which corresponds to the
multiplicity bins $M3$ and $M4$ as defined by the Plastic Ball
collaboration \cite{dos87} at BEVALAC/SIS energies. As in
\cite{sahu98} we have calculated the flow
\begin{equation}
\label{flo} F = \left \langle \frac{d (P_x/A)}{dy} \right \rangle |_{y=0}
\end{equation}
by fitting a  linear plus cubic term as a function of
the normalized rapidity $y = y_{cm}/y_{proj}$
for Au + Au systems at all energies.

In Fig.~3 the transverse flow (\ref{flo})  is displayed in
comparison to the data from Refs. \cite{FOPI,EOS,aE877,aE895} as
collected in Ref.~\cite{SFLOW} for Au + Au systems. The solid line
(RBUU with $\sqrt{s_{sw}}$ = 3.5 GeV) is obtained with the scalar
and vector self energies as discussed above, i.e. Eq.
(\ref{form}). The dotted line (CASCADE with $\sqrt{s_{sw}}$ = 3.5
GeV) corresponds to cascade calculations for reference in order to
show the effect of the mean field relative to that from
collisions. We observe that the solid line (RBUU, cf. Fig.~1) is
in good agreement with the flow data at all energies; above
bombarding energies of 6 A GeV the results are practically
indentical to the cascade calculations showing the potential
effects to become negligible.

The sideward flow shows a maximum around 2
$A$ GeV for Au + Au and decreases continuously at higher beam
energy ($\ge 2 A$GeV) without showing any explicit minimum
\cite{Rischke}. This is due to the fact that the repulsive force
caused by the vector mean field  decreases at high beam energies
(cf. Fig. 1) such that in the initial phase of the collision there
are no longer strong gradients of the potential within the
reaction plane. In subsequent collisions, which are important for
the Au + Au  due to the system size, the kinetic energy of the
particles relative to the local rest frame is then in a range
($E_{kin} \leq$ 1 GeV) where the Schr\"odinger equivalent
potential (at density $\rho_0$) is determined by the experimental
data~\cite{ham90}.

We thus conclude that for the sideward flow data up to $\le 11 A$GeV one
needs a considerably strong vector potential at low energy and that one has
to reduce the vector mean field at high beam energy roughly
in line with Fig. 1.
In other words, there is only a weak repulsive force at high relative
momenta and high densities.

Another aspect of the decreasing sideward flow can be related to
the dynamical change in the resonance/string degrees of freedom as
already discussed above. For instance, for $\sqrt{s_{sw}} = 2.6$
GeV the calculated flow turns out to be smaller than the data
above 1.5 $A$ GeV and approaches the cascade limit already for
$\approx$ 3-4 $A$ GeV. This is due to the fact that in strings the
incident energy is stored to a larger extent in their masses and
the translational energy is reduced accordingly. In addition -- in
the present treatment of string formation and decay -- strings do
not interact with other hadrons before their decay and thus do not
contribute to pressure during this time. In this respect string-hadron
or string-string interactions might become important when a
sufficient part of hadrons are excited to strings during the
heavy-ion reaction. We recall that the role of string fusion to create
multi-strange (anti-)hyperons has already been pointed out in
Refs.~\cite{soff99,Sorge}.

\subsection{Elliptic flow}
Apart from the directed flow of protons, the elliptic flow
provides additional information and constraints on the nuclear potentials
involved \cite{dani98}. In this respect the elliptic flow for protons
\cite{vol96}
\begin{equation}
v_2 = \left \langle \frac{(P_x^2-P_y^2)}{(P_x^2+P_y^2)} \right \rangle
\end{equation}
 for $|y_{cm}/y_{proj}| \leq 0.2 $ is shown in Fig. 4 as
a function of the incident energy for Au + Au collisions at $b$ =
6 fm. The solid line (RBUU with $\sqrt{s_{sw}}$ = 3.5 GeV) is
obtained with the same mean fields as in Fig. 3 while the dotted
line (CASCADE with $\sqrt{s_{sw}}$ = 3.5 GeV) stands again for the
cascade results. The flow parameter $v_2$ changes its sign from
negative at low energies ($\le 5 A$ GeV) to positive elliptic flow
at high energies ($\ge 5 A$ GeV). We observe that the solid line
is in good agreement with data, where the data points are from the
EOS \cite{EOS}, E895 \cite{aE895} and E877 \cite{aE877}
collaborations and taken from Ref. \cite{bE895}.

This can be understood as follows: At low energies the squeeze-out
of nuclear matter leads to a negative elliptic flow since
projectile and target spectators distort the collective expansion
of the 'fireball' in the reaction plane. At high energies the
projectile and target spectators do not hinder anymore the
in-plane expansion of the 'fireball' due to their high velocity
($\approx c$); the elliptic flow then is positive. The competition
between squeeze-out and in-plane elliptic flow at AGS energies
depends on the nature of the nuclear force as pointed out already
by Danielewicz et al. \cite{dani98}. We note, however, that in our
calculation with the momentum-dependent potential (Fig. 1) we can
describe both the sideward as well as elliptic flow data
\cite{FOPI,rei97,EOS,aE877,bE877,SFLOW,aE895,bE895} simultaneously.

In our cascade calculation (Fig. 4) the elliptic flow from
squeeze-out is weaker due to the lack of a nuclear force; this
demonstrates the relative role of the momentum-dependent nuclear
potentials on the $v_2$ observable below bombarding energies of about
5 A GeV.

\section{Summary}

In this work we have calculated the baryon sideward and elliptic
flow in the energy range up to 11 $A$ GeV in the relativistic
transport model RBUU for Au + Au collisions. We found that in
order to reach a consistent understanding of the nucleon optical
potential up to 1 GeV, the transverse mass distributions of
protons at AGS energies as well as the excitation
function of sideward and elliptic flow
\cite{FOPI,rei97,EOS,aE877,bE877,SFLOW,aE895,bE895} up to 11 A
GeV, the strength of the vector potential has to be reduced in the
RBUU model at high relative momenta and/or densities. Otherwise,
too much flow is generated in the early stages of the reaction and
cannot be reduced at later phases where the Schr\"odinger
equivalent potential is experimentally known. This constrains the
parameterizations of the explicit momentum dependence of the
vector and scalar mean fields $U^\nu$ and $U^S$ in Eq. (1) at high
momenta where no data for elastic $pA$ scattering are available.

In addition, we have shown the relative role of resonance and
string degrees of freedom at AGS energies. By reducing the number
of degrees of freedom via high mass resonances one can build up a
higher pressure and/or temperature of the 'fireball' which shows
up in the transverse mass spectra of protons as well as in the
sideward flow \cite{hom98}. A possible transition from resonance
to string degrees of freedom is indicated by our calculations at
invariant baryon-baryon collision energies of $\sqrt{s} \approx$
3.5 $\pm 0.3$ GeV which corresponds to a proton laboratory energy
of about 3.6 -- 5.8 GeV. Due to Fermi motion of the nucleons in Au
+ Au collisions the transition from resonance to string degrees of
freedom becomes smooth and starts from about 3 A GeV; at 11 A GeV
practically all initial baryon-baryon collisions end up in
strings, i.e. hadronic excitations in the continuum that decay to
hadrons on a time scale of about 0.8 fm/c in their rest frame.
This initial high-density {\it string matter} (up to 10 $\rho_0$
at 11 $A$ GeV) should not be interpreted as {\it hadronic matter}
since it implies roughly 5 constituent quarks per fm$^3$, which is
more than the average quark density in a nucleon.

\vspace{1cm} The authors are grateful to D. Keane for valuable
comments on the experimental data sets. One of us (PKS) likes to
acknowledge the support from the JSPS, Japan. This work was also
supported by GSI Darmstadt.

\newpage
\vskip 0.15in
\psfig{figure=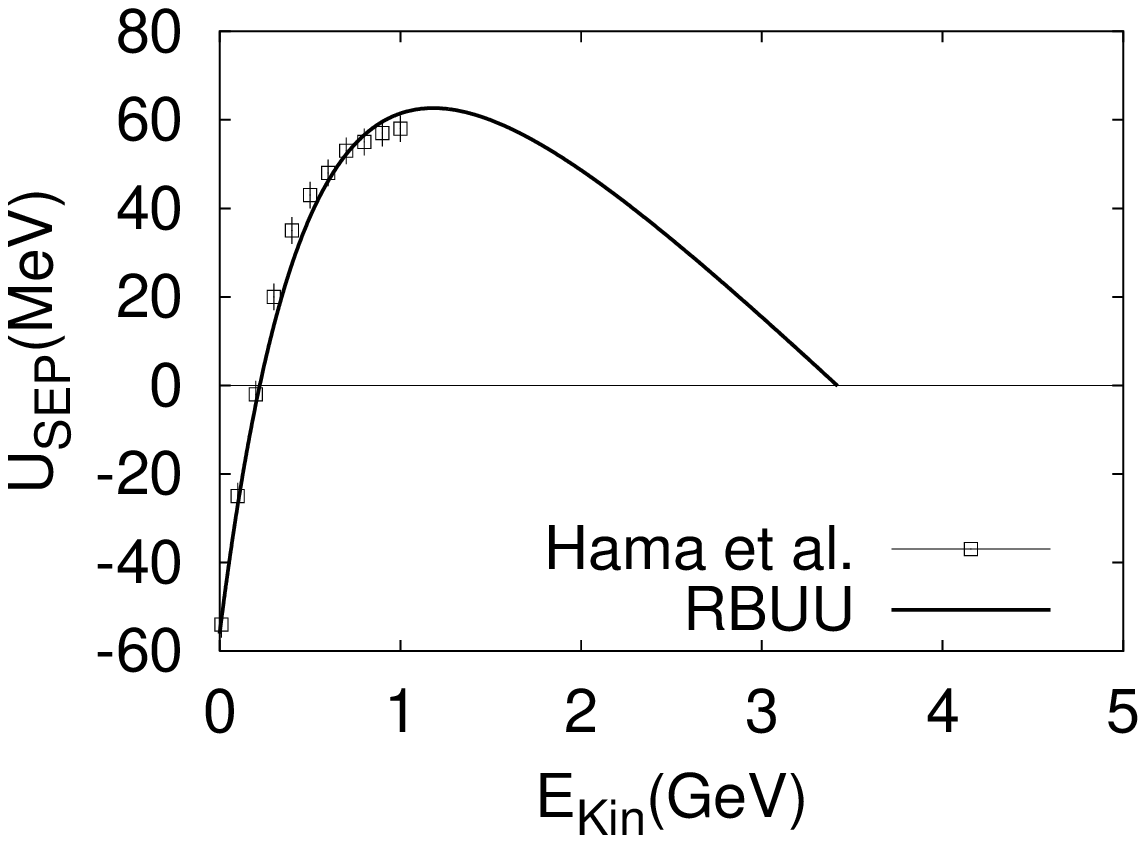,width=14cm}
\vskip 0.15in
{\noindent
\small {{\bf Fig.1} The Schr\"odinger equivalent potential
(\protect\ref{pot}) at density $\rho_0$ as a
function of the nucleon kinetic energy $E_{kin}$.
The solid curve (RBUU) results from the
momentum-dependent potentials discussed in the text.
The data points are from Hama et al. \protect\cite{ham90}.
\mbox{} }}

\newpage
\vskip -5.0cm \hskip 1.5cm \psfig{figure=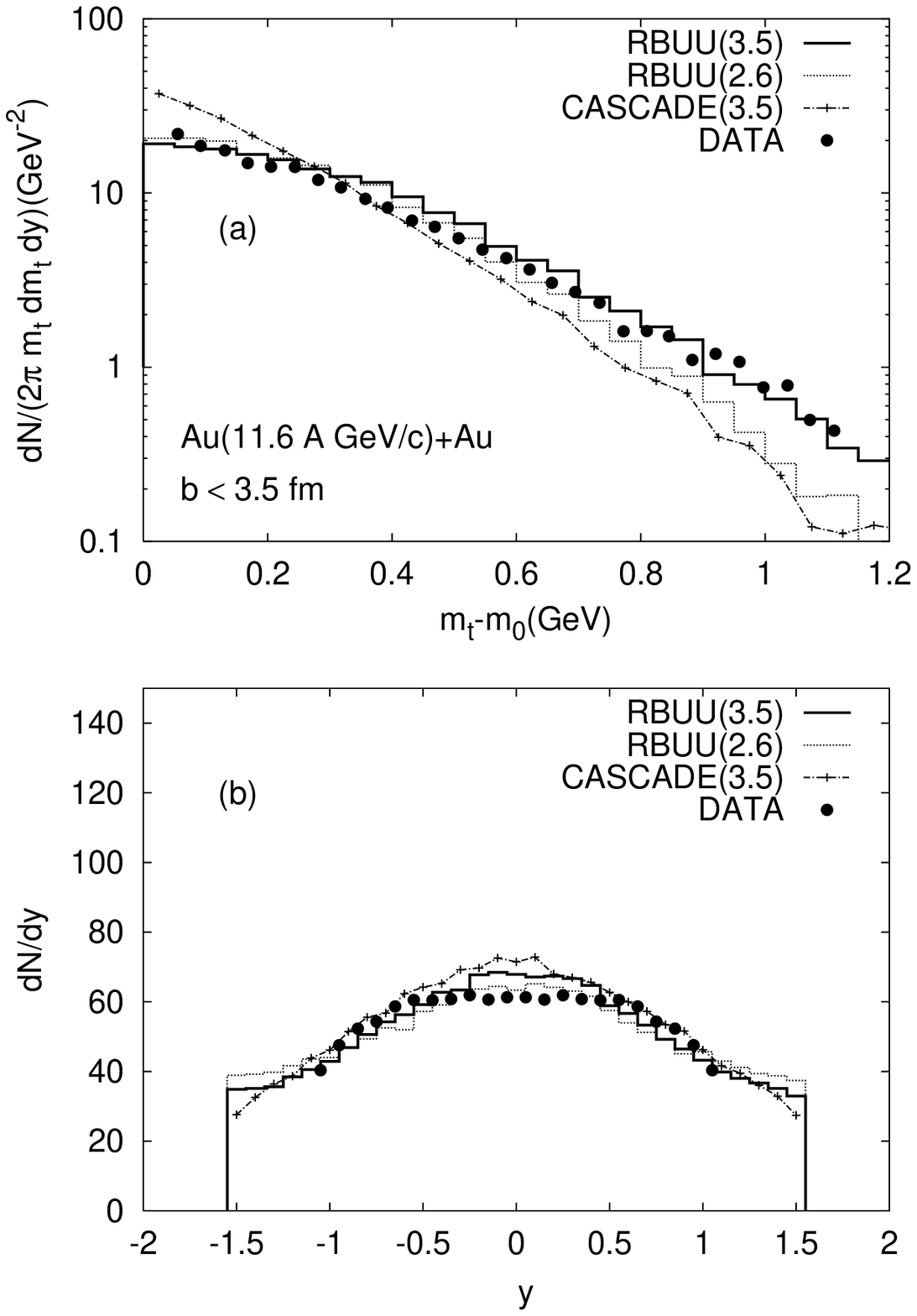,width=8cm}
\vskip 0.15in {\noindent \small {{\bf Fig.2} (a) The transverse
mass spectra of protons for Au + Au collisions at $b < 3.5$ fm.
The solid line and the dot- dashed line with crosses are results
for $\sqrt{s_{sw}}$=3.5 GeV with and without nuclear potentials,
respectively. The dotted line RBUU(2.6 GeV) is for
$\sqrt{s_{sw}}$=2.6 GeV. The data points are taken from the E802
collaboration \cite{E802}. (b) Same as (a) but for the proton
rapidity distribution.

\newpage
\vskip -0.15in \psfig{figure=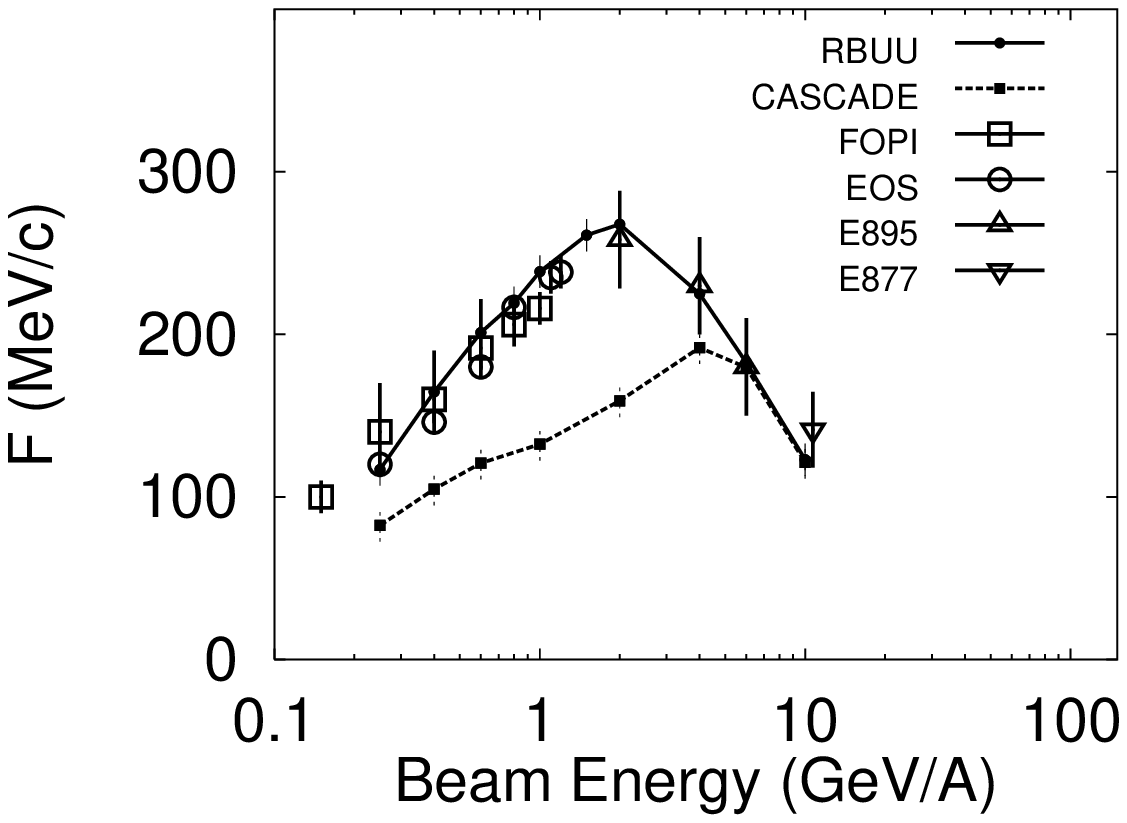,width=12cm} \vskip 0.15in
{\noindent \small {{\bf Fig.3} The sideward flow $F$ as a function
of the beam energy per nucleon for Au + Au collisions at $b=6$ fm
from the RBUU calculations. The solid line results for the
parameter set RBUU, the dotted line for a cascade calculation with
$\sqrt{s_{sw}}$ = 3.5 GeV. The data points are from the FOPI, EOS,
E895 and E877 collaborations \cite{FOPI,EOS,aE877,aE895,bE895}.}

\newpage
\vskip -0.15in \psfig{figure=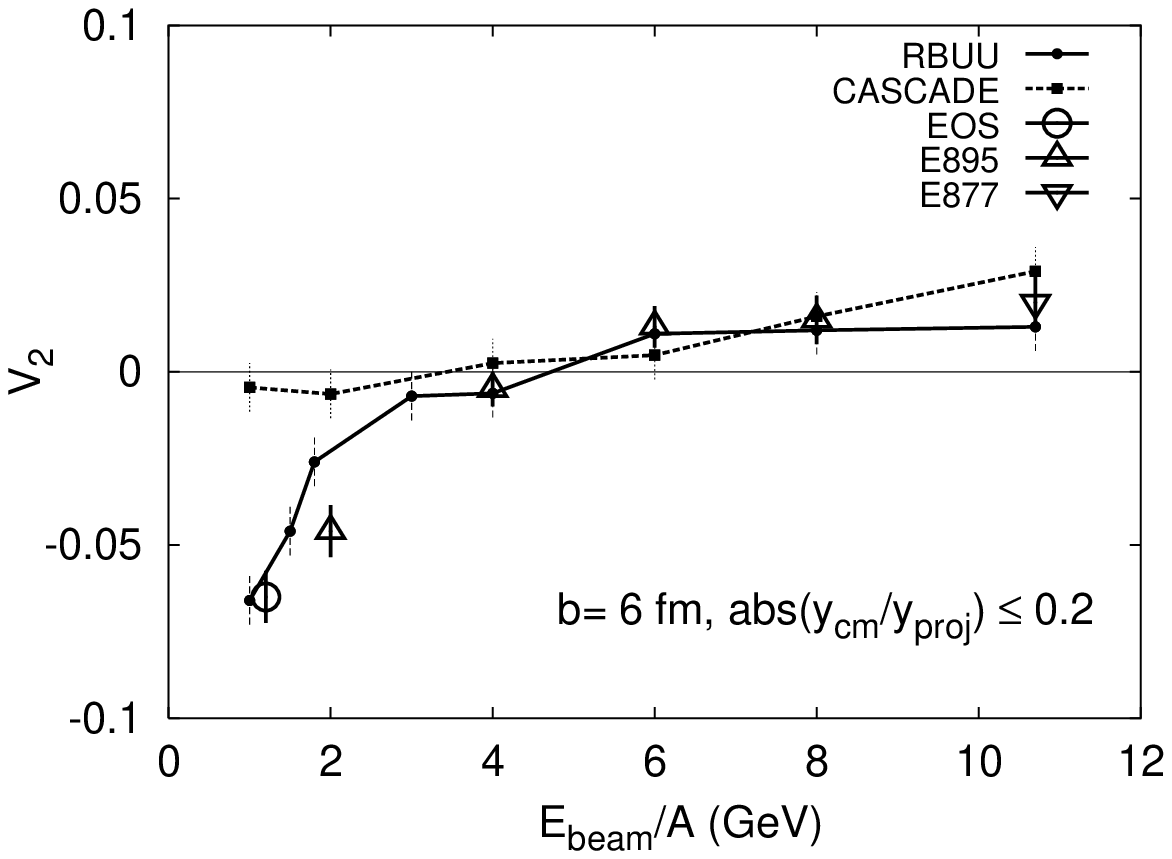,width=12cm} \vskip 0.15in
{\noindent \small {{\bf Fig.4}  The elliptic flow $v_2$ of protons
versus the beam energy per nucleon for Au + Au collisions at $b=6$
fm from the RBUU calculations. The solid line results for the
parameter set RBUU, the dotted line for a cascade calculation with
$\sqrt{s_{sw}}$ = 3.5 GeV. The data points are from the EOS, E895
and E877 collaborations \cite{EOS,aE877,aE895,bE895}. \mbox{} }}

\end{document}